# Using Blockchain to Achieve Decentralized Privacy In IoT Healthcare

Sajad Meisami
*Department of Electrical Engineering*
Sharif University of Technology
Tehran, Iran
meysami.sajad@ee.sharif.edu

Mohammad Beheshti-Atashgah
*Department of Electrical Engineering*
Sharif University of Technology
Tehran, Iran
m_beheshti_a@mut.ac.ir

Mohammad Reza Aref
*Department of Electrical Engineering*
Sharif University of Technology
Tehran, Iran
aref@sharif.edu

*Abstract—* With the advent of the Internet of Things (IoT), e-health has become one of the main topics of research. Due to the sensitivity of patient information, patient privacy seems challenging. Nowadays, patient data is usually stored in the cloud in healthcare programs, making it difficult for users to have enough control over their data. The recent increment in announced cases of security and surveillance breaches compromising patients' privacy call into question the conventional model, in which third-parties gather and control immense amounts of patients' Healthcare data. In this work, we try to resolve the issues mentioned above by using blockchain technology. We propose a blockchain-based protocol suitable for e-health applications that does not require trust in a third party and provides an efficient privacy-preserving access control mechanism. Transactions in our proposed system, unlike Bitcoin, are not entirely financial, and we do not use conventional methods for consensus operations in blockchain like Proof of Work (PoW). It is not suitable for IoT applications because IoT devices have resources-constraints. Usage of appropriate consensus method helps us to increase network security and efficiency, as well as reducing network cost, i.e., bandwidth and processor usage. Finally, we provide security and privacy analysis of our proposed protocol.

## I. Introduction

Many societies in the world are facing a significant increase in the number of medical patients, and access to primary doctors or medical staff is becoming more difficult for patients. Internet of Things (IoT) allows any device to connect to other devices and the Internet at any time and any place, and researchers hope that more than 75 billion devices will be able to connect to the Internet by 2025 [1]. Fortunately, e-health is one of the main applications of IoT, and it can help medical staff be able to treat more patients and provide more comfort and convenience for patients. Patients can stay connected with medical staff as required. It also reduces medical costs and improves the quality of care and treatment.

In recent years, the improvement of the IoT has allowed us to use electronic wearable devices to implement e-health. Wearable devices measure a patient's vital signs like heart rate, blood glucose, body temperature, blood pressure, etc. These wearable devices automatically collect patient health data and transfer it to central storage or cloud for further processing by medical staff to facilitate health monitoring, disease diagnosis, and treatment [2].

Patients' healthcare data are highly privacy-sensitive, and sharing of data may raise the risk of exposure. These data usually stored on a server and processed remotely, and this raises concerns regarding the privacy and confidentiality of patients' healthcare data. Also, several security attacks are possible in such cases; for example, an attacker can track patients' data on the Internet and modify or replace it with incorrect data in the data center, or an attacker can steal information from remote servers [3].

### A. Organization

Section II describes the related work involved. Section III discusses the challenges we solve in this paper. Sections IV provides an overview of the system model, whereas section V explains the network protocol in detail. Section VI discusses the security and privacy of the proposed model, and section VII provides a conclusion.

## II. Related Work

To address the privacy issue on personal data, researchers developed various methods. One of them is data anonymization that attempts to protect personally identifiable information. For example, in the k-anonymity method used in anonymous datasets, any necessary recorded data is indistinguishable from at least k−1 other important recorded data [4]. However, in recent research, it has been shown, anonymized datasets can be broken even with a little information (their anonymity disappears) [5]. There exist other privacy-preserving methods like differential privacy, that perturbs data, or adds noise to the computational process before sharing the data [6] and practices on creating noisy data or summarizing [7]. These methods are not efficient in healthcare applications where patients' original data are required to send to medical staff for medical treatments.

Attribute-based encryption (ABE) is a useful technology that can provide data privacy and fine-grained access control when users want to secretly share data stored in a third-party cloud server [8]. Almost all ABE schemes require a trusted private key generator (PKG) to set up the system and distribute for users the corresponding secret key [9]. However, in all ABE schemes, the PKG can decrypt all data stored in the cloud server, which may cause serious problems such as privacy data leakage and key abuse. Furthermore, the traditional cloud storage model runs in a centralized storage manner, so existence a single point of failure can collapse the system. Other encryption schemes exist that allow running computations and queries over encrypted data that called fully homomorphic encryption (FHE) [10] but are currently too unsuitable to use in practice widely.

In recent years, decentralized cryptocurrency systems have emerged. Bitcoin was the first of these systems, which use blockchain technology. Bitcoin allows users securely to make transactions and transfer currency (bitcoins) with others without the need to trusted third-party [11]. Blockchain works as an immutable timestamp ledger of blocks that are shared across all participating nodes in the network, which can bypass the need for a central authority [12]. This technology is used for sharing and storing data in a distributed manner by a peer-to-peer network [13]. Nowadays, blockchain is playing an effective role in financial transactions [14]. Also, it can be a facilitator in many other fields. Such as decentralized IoT [15], identity-based PKI [16], decentralized supply chain [17], decentralized proof of document existence [18], decentralized storage [19–21], etc.

In [22], the work by Zyskind et al. has shown the use of blockchain technology to construct an access control and management platform for personal data. They focused on users' privacy. And they combine blockchain and off-chain storage to storing encrypted data out of blockchain ledger while pointers to the encrypted data exist on the blockchain. Recently researchers addressed the security and privacy issues on healthcare data, using blockchain technology, and proposed new schemes [23–26].

In this paper, we combine IoT and blockchain technology to construct a novel platform for patients' healthcare data management that satisfies privacy and security requirements.

### III. CHALLENGES AND OUR SOLUTIONS

Secure transmission and preservation of privacy for the patients' healthcare data are the main concerns in the e-health application of IoT. However, the decentralized essence of the blockchain and other attributes like immutability and transparency make blockchain very suitable for e-health. But there are still challenges for applying blockchain into the IoT. Now we discuss the challenges and explain our solutions.

#### A. Scalability

IoT devices are resource-restricted, so it's hard for them to solve computationally intensive problems to execute a consensus algorithm for adding a new blocks to the blockchain ledger, especially in Proof-of-Works (PoW). We eliminate the concept of PoW for consensus in our blockchain network and use a method called Practical Byzantine Fault Tolerance (PBFT), a consensus method based on voting [27]. BPFT involves multiple rounds of voting by all nodes of the network [28]. BPFT helps us to reduce network costs, i.e., bandwidth, need for processors, and energy for consensus operations.

#### B. Data Storage

It's not practical and suitable to store IoT big data on the blockchain ledger. Because of this, we do not store the data on the blockchain, but only we store the pointers to the data (hash of encrypted data) on blockchain to lighten the storage space of blockchain. The data (encrypted data) are stored on Off-chain storage (see section IV, E. Off-chain Storage).

#### C. Security of Data

Patients' healthcare data are highly sensitive. To satisfy the security of data, we use a symmetric key encryption scheme (see section V, A. Cryptographic Techniques Used in The System). At first, The data is encrypted by symmetric key encryption and then sent to the blockchain network. So even in Off-chain Storage, the information is stored in the form of encrypted data.

#### D. Patients' Privacy

The main concern that is addressed in this paper is preserving patients' privacy because Patients' healthcare data is highly privacy-sensitive. We assume that the medical staff is honest-but-curious (i.e., they follow the protocol). In our system, patients could be remain (pseudo) anonymous. At the same time, medical staff profiles could be stored on the blockchain so that patients can trust medical staff by verifying medical staff identities. Our proposed platform satisfies the following Items:

*1) Patients' Data Ownership:* In our platform, patients are the only owner of their healthcare data, and only they can control that data. As such, the platform recognizes the medical staff as Service Providers with granted permissions set and the patients as healthcare data owners.

*2) Fine-grained Access Control:* Each patient can grant a set of permissions to Any desired member of the medical staff for accessing a patient's healthcare data. Also, the patient can alter or revoke the set of granted access permissions. These permissions are securely stored on blockchain ledger as access-control policies, where only the patient can change or revoke them.

*3) Data Transparency and Auditability:* Patients have complete and Accurate transparency over their collected healthcare data, and they can know how medical staff can access to healthcare data.

### IV. OUR SYSTEM MODEL

As demonstrated in Figure 1, our proposed system model consists of the following main modules:

1) Wearable IoT devices.
2) Patient's smartphone.
3) Medical staff.
4) Blockchain.
5) Off-chain storage.

#### A. Wearable IoT devices

The wearable IoT devices will collect all healthcare data from the patient's body, such as blood pressure, body temperature, and heart rate. These devices have limitations in resources such as low energy, low processing power, and little storage space. By considering these limitations, wearable IoT devices send the collected healthcare data to the patient's smartphone using short-range communication, such as Bluetooth or Zigbee.

#### B. Patient's Smartphone

Smartphones are more potent than IoT devices because smartphones have more storing space, more battery life, and higher processing power. So smartphones can do more complicated works, such as computational and cryptographic operations. Also, smartphones can transmit data via long-range communications (such as cellular networks) that can play the role of a gateway and allow patients to interact with the blockchain network.

#### C. Medical staff

Medical staff, including physicians and nurses, should receive patients' healthcare data and, after analyzing it, obtain information about the patient's health status. Then they provide appropriate treatment for patients.

#### D. Blockchain

In our work, we use blockchain to store access policies and eliminate the need for a third-party that preserves network against DoS attack and single point of failure. It also ensures the availability and integrity of the patients' data.

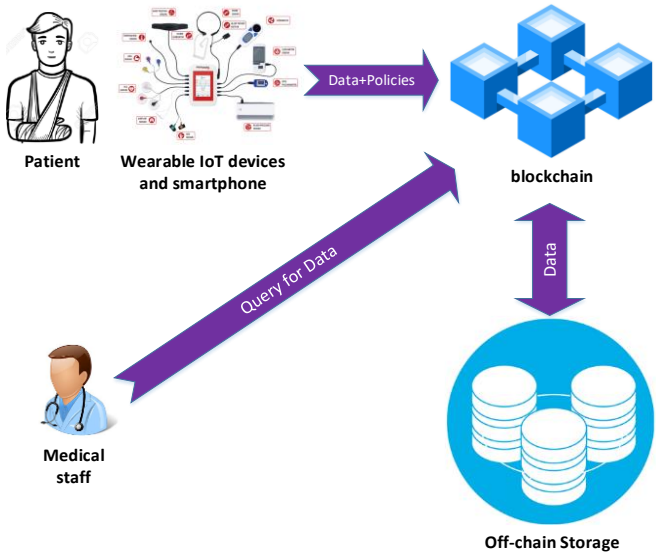

Fig. 1. Our proposed system model

The data is not stored on the blockchain, but only the pointers to the data (hash of encrypted data) are stored on it to lighten the storage space of blockchain. Also, because the use of PoW in IoT applications is not appropriate, we use PBFT for consensus operations.

*E. Off-chain Storage*

We store patients' encrypted data on off-chain storage. For the implementation of off-chain storage, we use The InterPlanetary File System (IPFS) [21] that is a peer-to-peer distributed file system that seeks to connect all computing devices with the same system of files. It provides a high throughput content-addressed block storage model, with content-addressed hyperlinks. IPFS combines a distributed hashtable (or DHT), an incentivized block exchange, and a self-certifying namespace. IPFS has no single point of failure, and nodes do not need to trust each other [21]. In IPFS, we distribute the data and store them on different servers all over the world. Not using the central server is the reason for the advantage of IPFS over conventional cloud storage.

## V. THE NETWORK PROTOCOL

In this section, we describe the protocol used in our system in detail.

*A. Cryptographic Techniques Used in The System*

*a) Hash Function:* We use SHA-256 [29] to implement the hash function ($H$ indicates the hash function).

*b) Symmetric Key Encryption:* Symmetric algorithm uses the same key for encryption of plaintext and decryption of ciphertext. We use AES [30] to implement the symmetric key encryption ($G_{enc}$ indicates generating algorithm).

*c) Digital Signature Scheme:* Digital signature is added to the data for authentication purposes. For the implementation of the digital signature, we use ECDSA with a secp256k1 curve [31] ($G_{sig}$ indicates generating algorithm).

*B. Protocol in Detail*

As illustrated in protocol 1, initially, patient $p$ and medical staff $m$ each generates a pair of private and public keys to sign and send transactions to the blockchain network and announce their public key (as their address) on the network. Patient $P$ also generates a secret key for encrypting data with an AES symmetric encryption algorithm. Then the patient shares the secret key with the chosen member of the medical staff so that, later, she will be able to decrypt her authorized data with that secret key.

| **Protocol 1** Joining the Blockchain |
|---|
| 1: **procedure** GENERATING($p,m$) |
| 2:     $p$ executes: |
| 3:         ($pk^p_{sig}$, $sk^p_{sig}$) ← $G_{sig}$( ) |
| 4:         $sk^{p,m}_{enc}$ ← $G_{enc}$( ) |
| 5:     $p$ shares $pk^p_{sig}$ (as address) on the network |
| 6:     $m$ executes: |
| 7:         ($pk^m_{sig}$, $sk^m_{sig}$) ← $G_{sig}$( ) |
| 8:     $m$ shares $pk^m_{sig}$ (as address) on the network |
| 9:     $p$ shares $sk^{p,m}_{enc}$ with $m$ from secure channel |
| 10:    // Both $p$ and $m$ have $sk^{p,m}_{enc}$ |
| 11:    **return** $pk^p_{sig}$, $pk^m_{sig}$, $sk^{p,m}_{enc}$ |
| 12: **end procedure** |

*a) Registration of access policy:* We denote the data access permissions by $POLICY_{p,m}$, which indicates the permissions that the patient $p$ gives to the selected member of the medical staff $m$ so that she can access a particular type or all of the patient's data. For example, $POLICY_{p,m}$ = *{body temperature, blood pressure}*.

By sending a $T_{access}$ transaction on the blockchain network that contains $POLICY_{p,m}$, the patient gives the desired permissions to the medical staff. As illustrated in protocol 2, this transaction is performed in the nodes of the blockchain, and it is checked the patient himself has sent the transaction, then the set of permissions are stored on the blockchain ledger.

The patient can send new $T_{access}$ transactions and change the permissions set granted to the medical staff. Also, Sending the empty set by the patient can revoke all access-permissions set previously granted.

| **Protocol 2** Access Control Transaction |
|---|
| 1: **procedure** ACCESSTX($pk^k_{sig}$ ,*message*) |
| 2:     $s ← 0$ |
| 3:     ($pk^p_{sig}$ // $pk^m_{sig}$ // $POLICY_{p,m}$ ) = *message* |
| 4:     **if** $pk^k_{sig}$ = $pk^p_{sig}$ **then** |
| 5:         $L[H(pk^k_{sig})]$ ← $L[H(pk^k_{sig})]$ ∪ *message* |
| 6:         // L is Blockchain memory |
| 7:         $s ← 1$ |
| 8:     **end if** |
| 9:     **return** $s$ |
| 10: **end procedure** |

TABLE I. COMPARISON OF DIFFERENT EXISTING SYSTEMS

| Model Name / Property | Yang[33] | Xia-I[34] | Xia-II[35] | Peterson[36] | Zang[37] | A.Zhang[38] | Our Proposed Model |
|---|---|---|---|---|---|---|---|
| **Access control** | ✓ | ✓ | ✓ | ✓ | ✓ | ✓ | ✓ |
| **Blockchain-Based** | × | ✓ | ✓ | ✓ | × | ✓ | ✓ |
| **Privacy Preserving** | ✓ | ✓ | ✓ | ✓ | ✓ | ✓ | ✓ |
| **IPFS Off-chain storage** | × | × | × | × | × | × | ✓ |

*b) Data storage and retrieval:* $T_{data}$ transaction is used to store patients' encrypted healthcare data on off-chain storage (IPFS) or access stored data and receive it. The patient (to store and retrieve the data) and the medical staff (only to retrieve the data) can send the $T_{data}$ transaction to the blockchain network.

If the $T_{data}$ transaction (by the patient or the medical staff) is sent to the network, the nodes in the blockchain first check with the following protocol (protocol3) whether they have access permissions or not?

---

**Protocol 3** Blockchain Permissions Checking
1: **procedure** POLICYCHECK($pk^k_{sig}$, T)    //T=type of data
2:     $s \leftarrow 0$
3:     **if** $L[H(pk^k_{sig})] \neq \emptyset$ **then**
4:         $(pk^p_{sig} || pk^m_{sig} || POLICY_{p,m}) \leftarrow L[H(pk^k_{sig})]$
5:         **if** $pk^k_{sig}=pk^p_{sig}$ **or**
6:             ($pk^k_{sig}=pk^m_{sig}$ and $T \in POLICY_{p,m}$) **then**
7:                 $s \leftarrow 1$
8:         **end if**
9:     **end if**
10:    **return** $s$
11: **end procedure**

---

Now, after checking the access permissions and the transaction sender's approval, he or she can store or retrieve patients' encrypted healthcare data with the following protocol.

---

**Protocol 4** Data Transaction
1: **procedure** DATATX($pk^k_{sig}$, message)
2:     $(C || T || RW) = message$
3:     // C=encrypted data, T=type of data
4:     //RW=read data(=1) or write data(=0)
5:     **if** POLICYCHECK($pk^k_{sig}$, T)=True **then**
6:         $(pk^p_{sig} || pk^m_{sig} || POLICY_{p,m}) \leftarrow L[H(pk^k_{sig})]$
7:         **if** $RW = 0$ **then**
8:             $L[pk^k_{sig} || T] \leftarrow L[pk^k_{sig} || T] \cup H(C)$
9:             (IPFS) $ds[H(C)] \leftarrow C$
10:            **return** $H(C)$
11:        **else if** $C \in L[pk^k_{sig} || T]$ **then**
12:            (IPFS) **return** $ds[H(C)]$
13:        **end if**
14:    **end if**
15:    **return** $\emptyset$
16: **end procedure**

---

Note that we used *IPFS* shorthand notation in lines 9 and 12 of Protocol 4 for accessing the off-chain storage. *IPFS* instruction cause to send Off-blockchain network message in off-chain storage for storing or retrieving data.

With the above protocols, the patient can easily upload the encrypted healthcare data in the network. The chosen members of the medical staff can also receive the encrypted data if there is a permission, and then with the $sk^{p,m}_{enc}$ that they have received before (in protocol 1), they can decrypt encrypted data and access the original healthcare data.

## VI. SECURITY AND PRIVACY ANALYSIS

In this section, we discuss and investigate the performance of our proposed protocol in terms of security and privacy. For security designing in any model, there are exist three main security requirements that need to be addressed: Confidentiality, Integrity, and Availability, known as CIA [32]. Confidentiality makes sure that the system's messages should be read by only authorized users who can access the system. The data integrity is responsible for making sure that no one without permission can change the stored data, and the availability of the data means that when users needed to the data, it is available to them. Now, we summarize the aforementioned primary security requirement evaluation in Table II.

TABLE II. SECURITY REQUIREMENT EVALUATION

| Requirement | Model Solution |
|---|---|
| Confidentiality | Achieved by using symmetric key encryption. |
| Integrity | Hashing of data blocks in blockchain is employed to achieve integrity. |
| Availability | Achieved by limiting acceptable transactions in the network. |
| Authorization | Using digital signature to achieve authorization. |

In our model, data ownership is emphasized for preserving privacy. That means only the patients (users) can control their data. The decentralized nature of the blockchain technology and using digital-signature to sign transactions in the network ensure that an adversary cannot be able to infiltrate the system as a user. Because gaining control over the majority of the network's resources (at least 51%), or forging digital-signature is almost impossible for the adversary. Also, the model ensures other privacy-preserving parameters that we previously mentioned (in section III) like, Data Transparency and Auditability. Fine-grained access control is satisfied by storing access-control policies on a blockchain ledger, where only the patient can change or revoke them.

In Table I, we have made a comparison between our proposed model and other existing systems. We considered a few attacks and analyzed the resilience of our model against each of them in Table III.

TABLE III. SECURITY ANALYSIS AGAINST ATTACKS

| Attack | Definition | Defence | Resilience |
|---|---|---|---|
| Denial of Service (DOS) Attack | Attacker generates a large number of transactions to increase traffic in the network and disrupt the blockchain. | Only two types of transactions can be sent in the network. Also, each node can send a limited number of transactions, and the blockchain network will reject the rest of the user's transactions after receiving a few messages from a specific node. | High |
| Modification Attack | Attacker modifies or removes the stored patients' data (like access policies and hash) on the blockchain ledger. | Blockchain uses an immutable ledger. | High |
| Public blockchain Modification | Attacker advertises a false ledger of blocks and makes it as the longest ledger. | We use the private type of blockchain, so the nodes are from outside the organization cannot work as miners to create a malicious block. | High |
| Storage Attack | Attacker wants to remove, change, or add data in the Off-chain storage. | On blockchain ledger exist a hash of the encrypted data stored in the Off-chain storage; therefore, changes in the data can easily be detected. | High |
| Appending Attack | Attacker compromises a miner and generates blocks with fake transactions to create a false reputation. | Due to the usage of private blockchain, so the users cannot generate a fake block, whereas miners in the blockchain must verify any transaction. | High |
| 51% Attack | Attacker controls more than 51% of miners and tries to compromise the consensus algorithm and generate a fake block. | The probability of occurrence of this attack is very low due to the usage of private blockchain and PBFT method for consensus. | High |
| Distributed DOS (DDOS) Attack | This is a distributed version of the Denial of Service (DOS) Attack. | A valid node can send a limited number of transactions in the network. After the blockchain network receives a transaction, miners check that received transaction has produced by a valid node then accept it. | Moderate |

## VII. CONCLUSION

Patients' healthcare data are privacy-sensitive and security-sensitive, and for managing them, we should not trust in the third-parties, where they are vulnerable to attacks and abuse. In this work, we proposed a novel platform based on the Internet of Things (IoT) and blockchain technology, motivated by the privacy and security challenges of patients' healthcare data in e-health. Our proposed platform enables patients to have ownership and full control over their sensitive healthcare data collected by their IoT wearable devices. This ownership and complete control over patients' data are satisfied by storing access control policies in a blockchain ledger by patients to specify who from medical staff can access patients' data. In this platform, we use off-blockchain storage to lightening the blockchain storage. Also, we use a suitable consensus method in the blockchain network due to the resource constraint factor of IoT.

ACKNOWLEDGMENT

The authors are thankful to Mr. Mohammad Doost and Mr. Mohammadhadi Ahmadiashtiyani for consulting and lengthy discussions on many of the ideas used in this paper.